\pdfoutput=1

\documentclass[11pt]{article}

\usepackage[final]{acl}

\usepackage{times}
\usepackage{latexsym}

\usepackage[T1]{fontenc}

\usepackage[utf8]{inputenc}

\usepackage{microtype}

\usepackage{inconsolata}

\usepackage{graphicx}

\usepackage{microtype}
\usepackage{inconsolata}

\usepackage{graphicx}
\usepackage{tabularx}
\usepackage{adjustbox}
\usepackage{booktabs}
\usepackage{multirow}
\usepackage{amsmath}
\usepackage{cuted}
\usepackage{comment}
\usepackage{lipsum}
\usepackage{colortbl}
\usepackage{xcolor}
\usepackage{enumitem}
\usepackage{bbding}
\usepackage{subcaption}

\definecolor{cmzhao}{rgb}{0.1, 0.8, 0.1}

\title{Cross-Modal Learning for Music-to-Music-Video Description Generation}

\author{
    \textbf{Zhuoyuan Mao}$^1$ \hspace{1em} \textbf{Mengjie Zhao}$^1$ \hspace{1em} \textbf{Qiyu Wu}$^1$ \hspace{1em} \textbf{Zhi Zhong}$^1$ \\ \textbf{Wei-Hsiang Liao}$^2$ \hspace{1em} \textbf{Hiromi Wakaki}$^1$ \hspace{1em} \textbf{Yuki Mitsufuji}$^{1,2}$ \\
    $^1$Sony Group Corporation \hspace{1em} $^2$Sony AI \\
    \texttt{\{zhuoyuan.mao, mengjie.zhao, qiyu.wu, zhi.zhong} \\ \texttt{weihsiang.liao, hiromi.wakaki, yuhki.mitsufuji\}@sony.com}
}

\begin{document}
\maketitle
\begin{abstract}
Music-to-music-video generation is a challenging task due to the intrinsic differences between the music and video modalities. The advent of powerful text-to-video diffusion models has opened a promising pathway for music-video (MV) generation by first addressing the music-to-MV description task and subsequently leveraging these models for video generation. In this study, we focus on the MV description generation task and propose a comprehensive pipeline encompassing training data construction and multimodal model fine-tuning. We fine-tune existing pre-trained multimodal models on our newly constructed music-to-MV description dataset based on the Music4All dataset, which integrates both musical and visual information. Our experimental results demonstrate that music representations can be effectively mapped to textual domains, enabling the generation of meaningful MV description directly from music inputs. We also identify key components in the dataset construction pipeline that critically impact the quality of MV description and highlight specific musical attributes that warrant greater focus for improved MV description generation.
\end{abstract}

\section{Introduction}

\begin{figure*}[t]
    \centering
    \includegraphics[width=\linewidth]{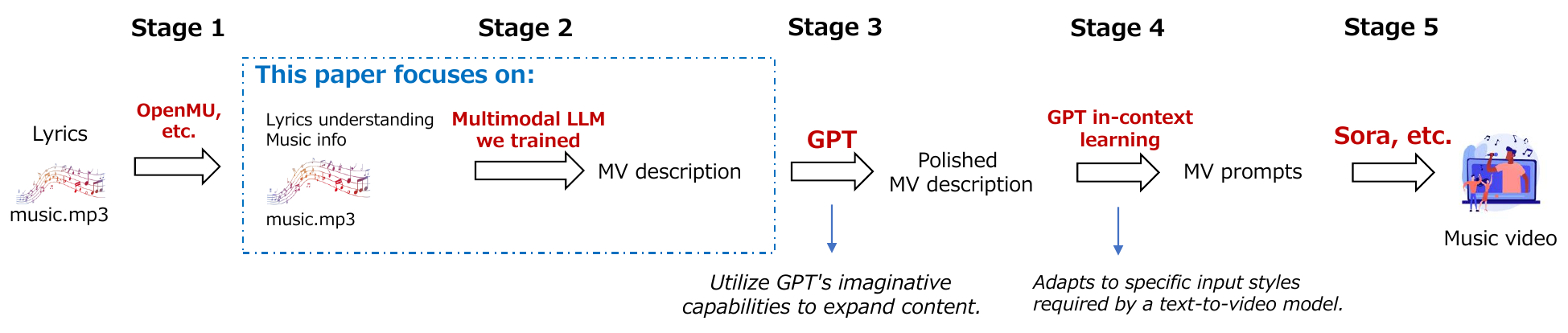}
    \caption{\textbf{Pipeline of music-to-MV generation.} We focus on multimodal model training of Stage 2 in this study.}
    \label{fig:pipeline}
\end{figure*}

Generating a music-video (MV) to match a given piece of music is a challenging task due to the inherent differences between the music and video modalities. Despite the challenges, MV generation holds significant potential for enhancing the music experience by providing a visual narrative that aligns with the music's tone, style, and mood, offering a more immersive and engaging way for audiences to connect with the music. Compared to generating music or audio from a given video~\cite{DBLP:journals/corr/abs-2406-04321,DBLP:journals/eswa/KangPH24}, the reverse task is more complex, as the video modality typically conveys richer spatial and temporal information than music. However, with the advent of text-to-video diffusion models~\cite{DBLP:journals/corr/abs-2408-06072,DBLP:journals/corr/abs-2410-13720,DBLP:journals/corr/abs-2412-03603}, videos can now be generated from textual descriptions. This development enables MV generation to be divided into two subtasks: (1) music-to-MV description generation and (2) MV description-to-MV generation. As illustrated in Fig.~\ref{fig:pipeline}, MV descriptions can be further refined using large language models (LLMs) like GPT~\cite{DBLP:journals/corr/abs-2303-08774} to fit specific text-to-video models~\cite{DBLP:conf/iccv/KhachatryanMTHW23}. \emph{In this study, we focus on the first task: generating MV descriptions from music}.

To this end, we propose a practical pipeline for data construction and model training to generate meaningful MV descriptions based on music inputs. Additionally, we explore methods to enhance the alignment of the generated descriptions to the given music. Specifically, we investigate the impact of various data sources—such as music, music genre tags, MV type tags, and lyrics understanding text—on the quality of the generated MV descriptions when fine-tuning multimodal LLMs like NExT-GPT~\cite{DBLP:conf/icml/Wu0Q0C24}. As shown in Fig.~\ref{fig:pipeline}, our approach first leverages existing music understanding models~\cite{DBLP:journals/corr/abs-2410-15573,DBLP:journals/corr/abs-2502-12623} to extract lyrics understanding text. We then fine-tune a multimodal LLM to process these diverse inputs and generate MV descriptions. The training data is constructed from gold-standard MVs, incorporating music-related information to enhance the connection between music and the generated descriptions. Unlike prior studies on MV generation, such as ViPE~\cite{shahmohammadi-etal-2023-vipe}, which focused solely on lyrics as input, our work emphasizes leveraging multiple modalities and evaluates the effectiveness of various combinations of input data in connecting multimodal representations for MV description generation.

To facilitate this study, we construct a music-to-MV description training and evaluation dataset using the Music4All dataset~\cite{DBLP:conf/iwssip/SantanaPDCMCFD20}. Empirical results on the NExT-GPT baseline and multimodal LLMs fine-tuned with our dataset demonstrate that meaningful MV descriptions can be generated from music and music-related textual inputs after multimodal fine-tuning. An ablation study on different combinations of input sources, including music, music genre tags, MV type tags, and lyrics understanding text, reveals that music and MV type tags are key components for high-quality MV description generation. While music genre tags and lyrics understanding text also contribute positively, they can be used interchangeably. Our findings can contribute to future study on enhancing MV descriptions and temporal alignment between music, lyrics, and the generated MV.

\begin{figure*}[t]
    \centering
    \includegraphics[width=0.9\linewidth]{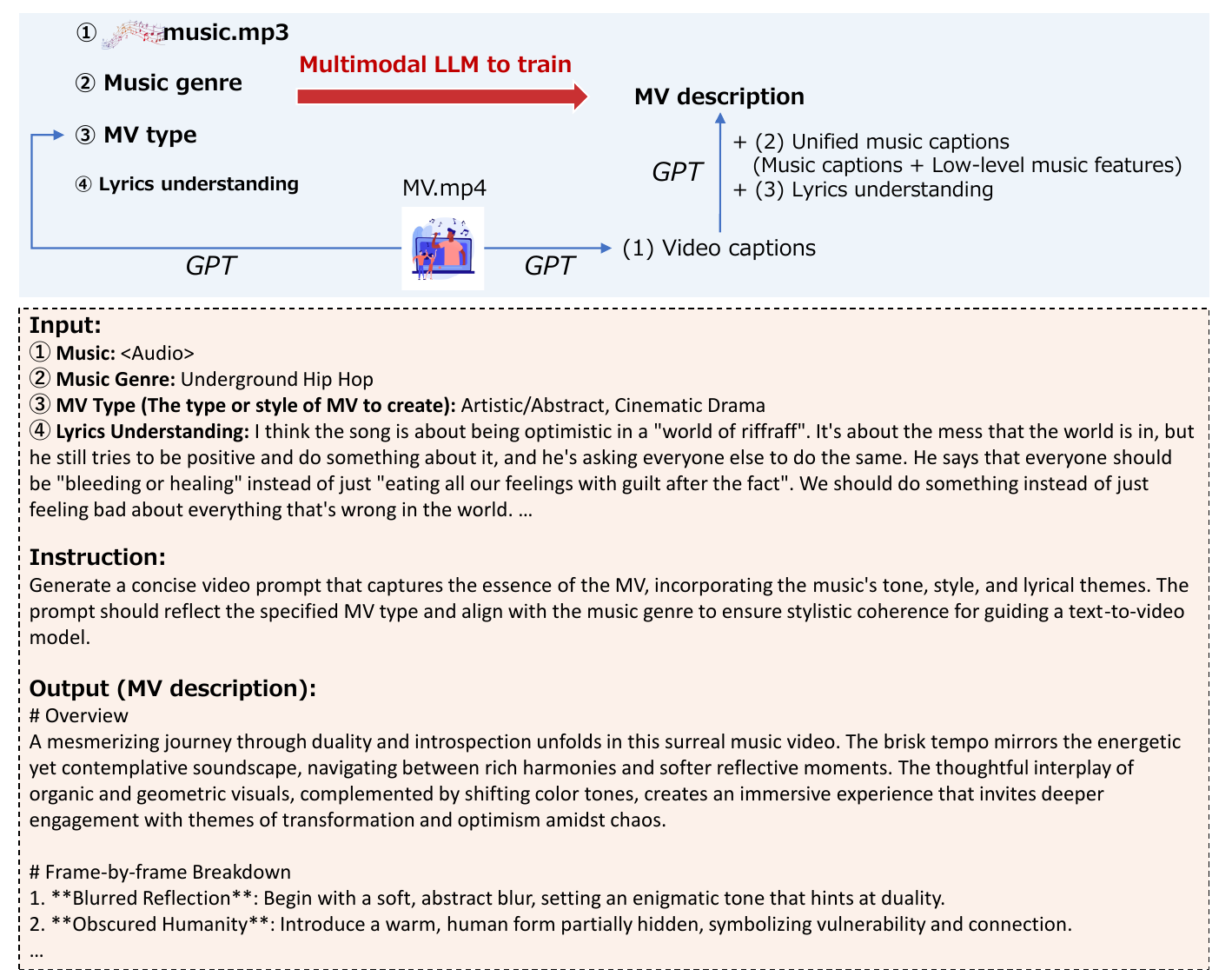}
    \caption{Process for creating music-to-MV description training datasets (top) and an example of utilizing the generated data to train music-to-text LLMs (bottom).}
    \label{fig:training}
\end{figure*}

\section{Proposed Method}
In this section, we present the pipeline proposed for training a multimodal LLM specifically tailored to the music-to-MV description generation task. For the first time, our pipeline incorporates a broader range of musical information beyond lyrics as inputs, aiming to enrich the holistic understanding of the music. Additionally, we introduce strategies to ensure the generated MV descriptions are more closely aligned with the musical inputs. The curated dataset is then utilized to fine-tune a multimodal LLM for performing the MV description generation task.

\subsection{Data Construction}
This section outlines our proposed pipeline for constructing training and evaluation datasets for the music-to-MV description generation task.

\subsubsection{MV Datasets}
We construct our datasets based on the Music4All dataset~\cite{DBLP:conf/iwssip/SantanaPDCMCFD20}, which comprises approximately 100k music clips paired with corresponding MVs and enriched with metadata such as energy, valence, and genre. To enhance the dataset, we leverage the OpenMU model~\cite{DBLP:journals/corr/abs-2410-15573} to generate lyrics understanding text for all music clips in Music4All. This process effectively interprets the lyrics for each piece of music, providing concise textual information related to the lyrics. Additionally, we filter out MVs that consist solely of static images rather than actual video footage. After filtering, the final dataset includes $56,446$ samples, $55,000$ for training and $1,446$ for testing.

\subsubsection{Construction of Input Data for Music and Associated Information}
After preparing the training and evaluation splits of music clips, MVs, lyrics understanding text, and metadata from the Music4All dataset, we curate various data types as inputs for the MV description generation task. To incorporate richer musical information across different modalities, we include music genre tags and lyrics understanding text as inputs in addition to the music clips. Moreover, to refine the output MV descriptions and make the task less open-ended, we specify the style of the output by providing MV type tags. These tags are assigned to the MV clips using GPT-4o mini~\cite{DBLP:journals/corr/abs-2303-08774} and include ten category candidates: Live Performance, Lyric Video, Animation, Story Narrative, Artistic/Abstract, Dance Performance, Behind-the-Scenes, Nature/Scenic, Static/Dynamic Picture Montage, and Cinematic Drama.\footnote{
Generated based on suggestions from GPT-4o mini.}

As shown in Fig.~\ref{fig:training}, the four types of inputs are used to train the multimodal LLM, guided by a fixed instruction:
``Generate a concise video prompt that captures the essence of the MV, incorporating the music's tone, style, and lyrical themes. The prompt should reflect the specified MV type and align with the music genre to ensure stylistic coherence for guiding a text-to-video model.''

\subsubsection{Construction of Output Data for MV Description}
The output MV descriptions should provide rich content detailing the visual elements of the MV while remaining closely tied to musical features, such as tempo, downbeats, and high-level characteristics like the mood conveyed by the music. To achieve this, we first utilize GPT-4o mini to caption MV clips and extract relevant visual contexts. Next, we refine these captions using GPT-4o mini again, integrating key musical features, including music captions, low-level musical attributes, and lyrics understanding. Music captions and lyrics understanding texts are generated using the OpenMU music understanding model, while low-level musical features are extracted with open-source tools~\cite{DBLP:conf/mm/BockKSKW16}, following the methodology of LLark~\cite{DBLP:conf/icml/GardnerDSB24}. The constructed MV description dataset includes two main components: an overview and a frame-by-frame breakdown, with frame captions extracted at two-second intervals for each 30-second MV clip. Examples of music captions, low-level music features, and a complete version of an MV description are provided in Appendix~\ref{sec:appendix-a}.

\subsection{Multimodal Model Training}
We utilize NExT-GPT~\cite{DBLP:conf/icml/Wu0Q0C24}, an any-to-any multimodal training framework, to fine-tune our model using the constructed music-to-MV description datasets. Following NExT-GPT's methodology, the fine-tuning process is divided into multiple stages. In the first stage, we fine-tune only the adaptor between the ImageBind~\cite{DBLP:conf/cvpr/GirdharELSAJM23} encoder and the Vicuna LLM~\cite{DBLP:conf/nips/ZhengC00WZL0LXZ23} utilizing the music captioning task. In the second stage, we simultaneously fine-tune the adaptor and apply LoRA~\cite{DBLP:conf/iclr/HuSWALWWC22} fine-tuning to Vicuna with the constructed music-to-MV description dataset. As illustrated in Fig.~\ref{fig:training}, the input data including the music clip is sequentially formatted, followed by a fixed instruction. The model is trained to generate MV descriptions comprising an overall summary and frame-by-frame breakdowns. We fine-tune for $5$ and $2$ epochs in the first and second stages, respectively, utilizing a learning rate of $1\mathrm{e}{-4}$ and a batch size of $2$. Training is conducted on $2$ NVIDIA A6000 GPUs. For LoRA, the rank and alpha are both set to $32$, following NExT-GPT.

\begin{table*}[t]
    \centering
    \resizebox{\linewidth}{!}{
    \begin{tabular}{lrrrrrrrr}
    \toprule
    \textbf{Model} & \textbf{BLEU-1} & \textbf{BLEU} & \textbf{ROUGE-P} & \textbf{ROUGE-R} & \textbf{ROUGE-F1} & \textbf{BERT-P} & \textbf{BERT-R} & \textbf{BERT-F1} \\
    \toprule
    \multicolumn{9}{l}{\textit{Baseline}} \\
    NExT-GPT~\cite{DBLP:conf/icml/Wu0Q0C24} & 8.3 & 0.2 & 20.7 & 9.2 & 11.8 & 80.9 & 76.5 & 78.6 \\
    \hline
    \multicolumn{9}{l}{\textit{Main results}} \\
    \textcircled{1}+\textcircled{2}+\textcircled{3}+\textcircled{4} & \textbf{42.9} & \textbf{14.6} & \textbf{22.9} & \textbf{23.2} & \textbf{22.7} & \textbf{87.4} & \textbf{86.4} & \textbf{86.9} \\
    \hline
    \multicolumn{9}{l}{\textit{Ablation study}} \\
    \textcircled{2}+\textcircled{3}+\textcircled{4} & 42.5 & 14.4 & 22.4 & 22.8 & 22.3 & 87.2 & 86.2 & 86.7 \\
    \textcircled{1}+\textcircled{2}+\textcircled{3} & \textbf{43.6} & \textbf{14.7} & \textbf{23.0} & \textbf{23.5} & \textbf{22.9} & \textbf{87.4} & \textbf{86.7} & \textbf{87.0} \\
    \textcircled{1}+\textcircled{3}+\textcircled{4} & \textbf{42.8} & \textbf{14.5} & \textbf{22.8} & \textbf{23.2} & \textbf{22.7} & \textbf{87.3} & \textbf{86.4} & \textbf{86.9} \\
    \textcircled{1}+\textcircled{2}+\textcircled{4} & 42.2 & 14.1 & 21.7 & 22.5 & 21.8 & 86.9 & 86.2 & 86.5 \\
    \textcircled{2}+\textcircled{3} & 41.8 & 14.0 & 21.8 & 22.4 & 21.8 & 87.2 & 86.1 & 86.6 \\
    \textcircled{1}+\textcircled{3} & 42.4 & 14.4 & 22.3 & 22.8 & 22.2 & 87.2 & 86.2 & 86.6 \\
    \textcircled{1}+\textcircled{4} & 41.3 & 13.8 & 21.4 & 22.4 & 21.6 & 86.8 & 86.0 & 86.4 \\
    \hline
    \multicolumn{9}{l}{\textit{Sanity check (w/o inputs, solely w/ instructions during inference)}} \\
    \textcircled{1}+\textcircled{2}+\textcircled{3}+\textcircled{4} & 39.3 & 13.2 & 20.2 & 22.5 & 21.0 & 85.8 & 85.6 & 85.7 \\
    \textcircled{1}+\textcircled{4} & 39.7 & 12.5 & 20.3 & 20.8 & 20.3 & 86.1 & 85.6 & 85.9 \\
    \bottomrule
    \end{tabular}
    }
    \caption{Results of MV description generation on the Music4All dataset. We provide BLEU-1 and BLEU-4 scores for BLEU, along with precision, recall, and F1 scores for both ROUGE-L and BERT-score. \textcircled{1}, \textcircled{2}, \textcircled{3}, and \textcircled{4} represent music, music genre tags, MV type tags, and lyrics understanding text, respectively. The top three values in each metric are highlighted in \textbf{bold}.}
    \label{tab:results}
\end{table*}

\section{Evaluation}
Using the $1,446$ test samples from our constructed dataset, we evaluate the generated MV descriptions with BLEU~\cite{papineni-etal-2002-bleu}, ROUGE-L~\cite{lin-2004-rouge}, and BERT-score~\cite{DBLP:conf/iclr/ZhangKWWA20}, considering different combinations of inputs: \textcircled{1} music, \textcircled{2} music genre tags, \textcircled{3} MV type tags, and \textcircled{4} lyrics understanding text. Additionally, we present several MV frames generated by Text2Video-Zero~\cite{DBLP:conf/iccv/KhachatryanMTHW23} to test the feasibility of the entire proposed MV generation pipeline in Appendix~\ref{sec:appendix-b}, using the ground-truth MV descriptions annotated by us as input. 

\subsection{Main Results}
As shown in Table~\ref{tab:results}, our proposed pipeline for music-to-MV description generation achieves significant improvements over the NExT-GPT baseline after fine-tuning for a specific music domain. This demonstrates that, with the proposed datasets and pipeline, music can be effectively mapped to the text domain. Comparing the main results with sanity checks that remove all inputs during inference (leaving only a fixed instruction), we observe that our carefully designed inputs for music-related information substantially contribute to the quality of the generated MV descriptions. Interestingly, after training, the model can generate reasonable MV descriptions even without any inputs, suggesting that the NExT-GPT model successfully adapts to the MV description generation downstream task.

\subsection{Ablation Study}
Through an ablation study exploring different combinations of data sources, we find that settings \textcircled{1}+\textcircled{2}+\textcircled{3} and \textcircled{1}+\textcircled{3}+\textcircled{4} achieve comparable or even slightly better performance to the full data combination (\textcircled{1}+\textcircled{2}+\textcircled{3}+\textcircled{4}). This suggests that the contributions of music genre tags (\textcircled{2}) and lyrics understanding text (\textcircled{4}) are interchangeable, without providing additional benefits when used together. Observing the results of \textcircled{1}+\textcircled{3}, we note that music genre tags (\textcircled{2}) and lyrics understanding (\textcircled{4}) positively impact the results and are not redundant inputs. When comparing the top three performing settings (\textcircled{1}+\textcircled{2}+\textcircled{3}, \textcircled{1}+\textcircled{3}+\textcircled{4}, and \textcircled{1}+\textcircled{2}+\textcircled{3}+\textcircled{4}) with the combinations \textcircled{2}+\textcircled{3}+\textcircled{4} and \textcircled{1}+\textcircled{2}+\textcircled{4}, we observe a significant performance drop. This highlights the importance of including both music (\textcircled{1}) and MV type tags (\textcircled{3}). Seeing the results of \textcircled{1}+\textcircled{4}, the simultaneous inclusion of music genre tags (\textcircled{2}) and MV type tags (\textcircled{3}) yields consistent improvement across all metrics. Moreover, the results of \textcircled{2}+\textcircled{3} demonstrate that even with simple tags for music and MV, the model can generate reasonable MV descriptions, suggesting future opportunities to enhance the model's performance by leveraging finer-grained features such as temporal alignment between lyrics and musical waves. 

\section{Conclusion}
In this study, we explored data construction and multimodal training pipelines for the music-to-MV description task, with the goal of building robust base models for the broader music-to-MV generation task. Our results on the constructed Music4All dataset highlight key data sources that significantly impact the quality of MV descriptions. Future work could extend our proposed dataset construction pipeline to additional music domains. Additionally, exploring more effective methods to align MV descriptions or MVs with the corresponding music could pave the way for improved performance in this challenging task.

\section*{Limitations}
The proposed approach has several limitations: (1) The pipeline was evaluated on a single constructed dataset. Testing on additional datasets could strengthen the claims made in this paper. (2) The pipeline focuses on converting music into MV descriptions for MV generation tasks, but relying solely on text descriptions may overlook important information necessary for effective MV generation. (3) Inputs were limited to music, music genre tags, MV type tags, and lyrics understanding text, while other features that could significantly enhance MV descriptions may not have been considered. (4) The data construction pipeline depends on LLMs for captioning, and the choice of LLMs could influence the quality of the generated MV descriptions.

\section*{Ethical Considerations}
In this study, we utilized only publicly available datasets and models to fine-tune the music-to-MV description generation task, ensuring no copyright issues. While our experiments focused on MV description generation, it is important to acknowledge that the fine-tuned models may produce potentially risky hallucinations. Users should use the generated content with caution, understand the risks associated with LLM-generated outputs, and implement content safety checks as post-processing. Although debiasing fine-tuning could help address these issues, it falls outside the scope of this work. Additionally, caution is needed when using text-to-video models based on the generated MV descriptions, ensuring that no illegal content, such as unauthorized human identities or privacy violations, is included.


\bibliography{anthology,custom}

\clearpage
\appendix

\begin{figure*}[t!]
    \includegraphics[width=\linewidth]{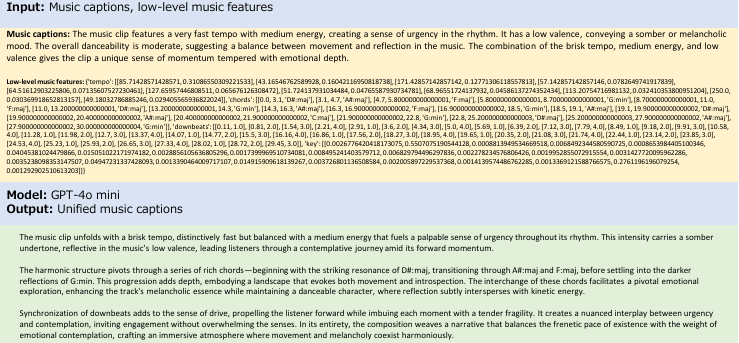}
    \caption{An example of music caption, low-level features and generated unified music captions.}
    \label{fig:mc}
\end{figure*}

\begin{figure*}[t!]
    \includegraphics[width=\linewidth]{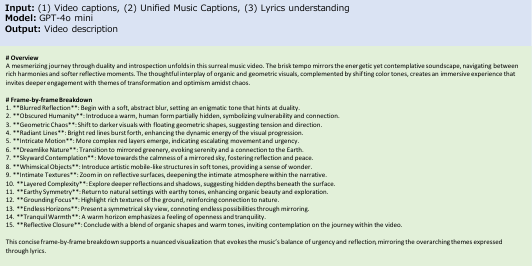}
    \caption{An example of a full MV description.}
    \label{fig:vd}
\end{figure*}

\section{Examples of How to Construct MV Description}
\label{sec:appendix-a}

We first extract low-level music features, including tempo, key, downbeats, and chords, using the open-source tool madmom~\cite{DBLP:conf/mm/BockKSKW16}. Based on these features and the textual captions of each music piece, we employ GPT-4o mini\footnote{\url{https://platform.openai.com/docs/models##gpt-4o-mini}} to generate unified music captions that seamlessly integrate all the musical information into natural, coherent sentences, as illustrated in Figure~\ref{fig:mc}. Subsequently, we prompt GPT-4o mini again to construct MV descriptions by combining the video captions of each gold-standard MV, the unified music captions, and the lyrics understanding text (see Figure~\ref{fig:training}). The resulting MV descriptions incorporate both visual and musical content, making them better suited for reconstructing the original MV.

\begin{figure*}[t!]
    \centering
    \includegraphics[width=\linewidth]{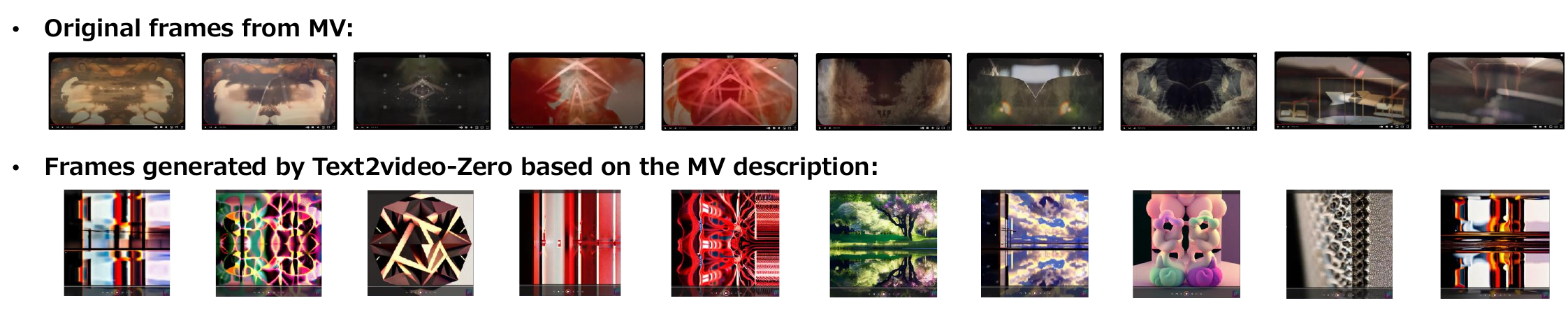}
    \caption{Frames from the original MV and generated by the Text2Video-Zero~\cite{DBLP:conf/iccv/KhachatryanMTHW23} model.}
    \label{fig:t2v}
\end{figure*}

\section{Generating Video Frames using MV Description}
\label{sec:appendix-b}
Figure~\ref{fig:t2v} showcases frames generated by the Text2Video-Zero~\cite{DBLP:conf/iccv/KhachatryanMTHW23} model using the gold-standard MV description example provided in Figure~\ref{fig:vd}. When compared to the original MV frames, we observe that even with only textual descriptions, the text-to-video model can produce content closely aligned with the intended visuals, such as the abstract geometric shapes in frames \#3 to \#5 and the mirrored sky in frames \#6 and \#7. This demonstrates the feasibility of our proposed pipeline for MV generation, as illustrated in Figure~\ref{fig:pipeline}. However, challenges remain, particularly in accurately generating complex elements like multi-layered imagery and human faces using current text-to-video models. Addressing these limitations will be crucial for future advancements in this domain.

\end{document}